\documentclass[11pt,paper]{article}
\usepackage{jheppub}

\usepackage{float,slashed}
\usepackage[usenames,dvipsnames,table]{xcolor}
\usepackage{graphicx,amsmath,amssymb,multirow,array,bm,mathrsfs}
\usepackage{epsf,amsfonts,slashed}
\usepackage[numbers,sort&compress]{natbib}
\usepackage[export]{adjustbox}
\usepackage{tikz}
\usepackage{scalerel,stackengine} 

\stackMath
\newcommand\reallywidehat[1]{%
\savestack{\tmpbox}{\stretchto{%
  \scaleto{%
    \scalerel*[\widthof{\ensuremath{#1}}]{\kern-.6pt\bigwedge\kern-.6pt}%
    {\rule[-\textheight/2]{1ex}{\textheight}}
  }{\textheight}%
}{0.5ex}}%
\stackon[1pt]{#1}{\tmpbox}%
}

\newcommand{\AY}[1]{{\textcolor{Red}{ [AY:#1] }}}

\newcommand{\BM}[1]{{\textcolor{Green}{ [BM:#1] }}}

\def\charge{\mu}

\title{
Multiply charged magnetic  black branes
}
\author{Ben~Meiring,}
\author{Ido~Shyovitz,}
\author{Sebastian~Waeber,}
\author{Amos~Yarom.}

\affiliation{Technion, Technion city, Haifa 32000, Israel}

\emailAdd{benwmeiring@gmail.com}
\emailAdd{ido.shyovitz@campus.technion.ac.il}
\emailAdd{s.f.d.waeber@gmail.com}
\emailAdd{ayarom@physics.technion.ac.il}

\abstract{
We discuss analytic solutions describing magnetically charged black branes in $d$ dimensional AdS space. Focusing on $d=5$, we study the response of the brane to an external short lived electric field. We argue that when the theory possesses an 't Hooft anomaly then at sufficiently low temperature a long lived oscillatory current will be observed long after the electric field has been turned off. We demonstrate this ``anomalous resonance'' effect via a numerical study.
}
\begin{document}
%


\maketitle

\section{Introduction}
The holographic duality \cite{Maldacena:1997re} has been found to be a useful tool in studying a variety of physical systems including heavy ion collisions \cite{Policastro:2001yc}, anomalous hydrodynamics \cite{Erdmenger:2008rm}, superfluidity and superconductivity \cite{Gubser:2008px,Hartnoll:2008vx,Hartnoll:2008kx,Herzog:2008he}, out of equilibrium steady states \cite{Chang:2013gba,Bhaseen:2013ypa,Lucas:2015hnv,Spillane:2015daa,Herzog:2016hob}, and turbulence \cite{Adams:2013vsa,Marjieh:2021gln,Waeber:2021xba} to name a few. More recently, there has been increased interest in using the duality to study thermal states in the presence of strong (external) magnetic fields \cite{DHoker:2009ixq,Kim:2010pu,Almuhairi:2010rb,DHoker:2010xwl,Ammon:2011je,Almuhairi:2011ws,Donos:2011qt,Donos:2011pn,Donos:2015bxe,Ammon:2017ded,Grozdanov:2017kyl,Landsteiner:2017hye,Landsteiner:2017lwm,Haack:2018ztx,Abbasi:2018qzw,Abbasi:2019rhy,Fernandez-Pendas:2019rkh,An:2020tkn,Amoretti:2020mkp,Gursoy:2020kjd,Ammon:2020rvg,Ballon-Bayona:2022uyy,Losacco:2022lnz,Das:2022auy,Rai:2023nxe}.

Thermally equilibrated states in the presence of an external magnetic (and/or electric) field on $\mathbb{R}^{d-1,1}$ of theories with a holographic dual are equivalent to asymptotically anti de Sitter (AdS) magnetically (and/or electrically) charged black brane solutions to Einstein-Maxwell theory with a negative cosmological constant. The value of the magnetic (and/or electric) field on the asymptotic boundary of AdS space represents the external magnetic (and/or electric) field in the dual theory.


Often, it is convenient to have an analytic description of the black brane metric. This allows for a better understanding of its dynamics and its response to perturbations. While analytic expressions for magnetically (and electrically) charged black branes exist in $3+1$ dimensional spacetimes, their higher dimensional analog does not seem to be available. The reason for the absence of analytic solutions in higher dimensions is likely associated with the lack of symmetry: In $3+1$ dimensions a constant magnetic field will pierce the (planar) event horizon so that rotational invariance (in the spatial directions associated with the asymptotic boundary) is maintained. In higher dimensions, a non vanishing magnetic field will necessarily break rotation invariance.

With the above symmetry considerations in mind, an analytic expression for magnetically charged black brane configurations can be obtained by allowing for several $U(1)$ gauge fields, such that each of their (equal in magnitude) magnetic fields pierce the horizon on a different plane so that rotation invariance is retained. Explicitly, the equations of motion following from
\begin{equation}
\label{E:thetheory}
	S = \int \sqrt{-g} \left(R + \frac{d(d-1)}{2 L^2} - \frac{1}{4}\sum_{ j=i+1}^{d-1} \sum_{i=1}^{d-1}  F^{ij}{}_{\mu\nu}F^{ij\,\mu\nu} \right) d^{d+1}x
\end{equation}
have charged black brane solutions of the form
\begin{align}
\begin{split}
\label{E:thesol}
	ds^2 & = -\frac{r^2}{L^2} A(r) dt^2 +2 dt dr + \frac{r^2}{L^2} \sum_{n=1}^{d-1}(dx^n)^2 \\
	A^{ij} &= \frac{1}{L^2}\left( \mu^{ij} - \frac{q^{ij}}{r^{d-2}} \right) dt + \frac{B}{L^4} x^j dx^i 
\end{split}
\end{align}
where
\begin{equation}
	A(r) = 1+ \frac{M}{r^{d}} -\frac{d-2}{4(d-4)} \frac{B^2}{r^4} + \frac{d-2}{2(d-1)} \frac{Q^2}{r^{2 d-2}} 
\end{equation}
for $d \neq 4$ with $M \in \mathbb{R}$, $Q^2 = \sum_{i,j}(q^{ij})^2$ and $F^{ij} = d A^{ij}$. The Greek indices $\mu,\nu = 0,\ldots,d$ run over the spacetime coordinates and $i,j=1,\ldots d-1$ count gauge fields. The solution for $d=4$ (AdS${}_5$) has been obtained in \cite{Donos:2011qt} and will be discussed in detail in section \ref{S:MBB}. In the remainder of this work we will set $L=1$. 

Clearly, the number of gauge fields required to support \eqref{E:thesol} grows as $d^2$ and therefore seems unrelated to physical applications. Fortuitously, for $d=3$ we require only one gauge field and then \eqref{E:thesol} reproduces the well known magnetically charged AdS${}_4$ black brane solution \cite{Hartnoll:2007ai}. In four dimensions, $3$ gauge fields are required which is somewhat unrelated to the electromagnetic field whose behavior one often tries to capture using holographic duality. However, $\mathcal{N}=2$ five dimensional gauged supergravity does contain three Abelian gauge fields. In the context of $\mathcal{N}=4$ super Yang Mills theory these can be thought of as sources for the maximal Abelian subgroup of the $SO(6)$ R-symmetry. 
Indeed, in the remainder of this work we will focus on ($d=4$)  five dimensional magnetically charged black brane solutions of the above type.

One interesting characteristic of magnetically charged black branes is their quasi normal modes (QNM's). Similar to their non rotationally invariant counterpart \cite{Bu:2016oba,Bu:2016vum,Ammon:2017ded,Haack:2018ztx,Bu:2018psl,Bu:2019mow}, the magnetically charged black branes we study possses quasi normal modes which approach the real axis as the magnetic field becomes very large compared to the temperature (in the presence of a Chern-Simons term). Given the analytic control we have over the black hole solution we are able to compute these quasi normal modes at perturbatively small temperatures (or large magnetic fields). We do this in section \ref{S:QNM}. 

As discussed in \cite{Haack:2018ztx}, since the quasi normal modes are long lived, perturbing the black brane at an appropriate frequency will excite this QNM and its oscillations may then be observed long after the perturbation has been turned off. We refer to this effect as an anomalous resonance and demonstrate its manifestation explicitly in section \ref{S:Driving}.



\section{Magnetically charged black branes in AdS${}_5$}
\label{S:MBB}
Our starting point is the action
\begin{equation}
\label{E:ansatz}
	S = \frac{1}{16 \pi G_N}\int \sqrt{-g} \left( R + {12}  - \frac{1}{4} \delta_{ij} F^i{}_{\mu\nu}F^{j\,\mu\nu} + \lambda \epsilon^{\mu\nu\rho\sigma\tau} A^1{}_{\mu}F^2_{\nu\rho}F^3_{\sigma\tau} \right) d^5x\,,
\end{equation}
which is related to the STU ansatz for the $D=5$, $N=2$ gauged supergravity action \cite{Behrndt:1998jd} once $\lambda=1/4$.
The equations of motion resulting from \eqref{E:ansatz} are given by
\begin{align}
\begin{split}
\label{E:EOM}
	R_{\mu\nu} + \frac{4}{L^2} g_{\mu\nu} & = \sum_i \left(\frac{1}{2} F^{i}{}_{\mu\alpha}F^{i}{}_{\,\nu}{}^{\alpha} - \frac{1}{12} g_{\mu\nu} F^{i}{}_{\alpha\beta} F^{i\,\alpha\beta}\right)  \\
	\nabla_{\mu}F_{1}{}^{\mu\nu} &= -\lambda \epsilon^{\nu\alpha\beta\gamma\delta}F_{2\,\alpha\beta}F_{3\,\gamma\delta}
\end{split}
\end{align}
and even permutations of the last equation for $F_{2\,\mu\nu}$ and $F_{3\,\mu\nu}$.

The action \eqref{E:ansatz} possesses an $SO(3)_f$ flavor symmetry rotating the three $U(1)$ fields which is broken by the Chern-Simons term. The ansatz
\begin{align}
\begin{split}
\label{E:ansatzsolution}
	ds^2 &= - 2 A(r)dt^2 + \Sigma(r)^2 \left(dx^2+dy^2+dz^2 \right) + 2 dt dr\,,
	\\
	A^1 &= {a^1(r)} dt + {B y} dz\,,
	\qquad
	A^2 = {a^2(r)} dt + {B z} dx\,,
	\qquad
	A^3 = {a^3(r)} dt + {B x} dy\,.
\end{split}
\end{align}
preserves a diagonal $SO(3)_B$ subgroup of the $SO(3)_f$ symmetry and spatial rotations once $a^i=0$. That is, \eqref{E:ansatzsolution} (with $a^i=0$) is invariant under a simultaneous spatial rotation in the $x$, $y$, $z$ directions, $R^{\mu}{}_{\nu}(\theta,\phi)$, and a rotation of the three $U(1)$ fields, $R^{i}{}_{j}(\theta,\phi)$, viz. $g_{\mu\nu} = g_{\alpha\beta}R^{\alpha}_{\mu}R^{\beta}_{\nu}$ and $F^i_{\mu\nu} = F^j_{\alpha\beta} R^{i}{}_{j} R^{\alpha}{}_{\mu}R^{\beta}{}_{\nu}$. Note that this ansatz is also invariant under parity in the $x^i$ coordinates.

After fixing the residual $r \to r+c$ symmetry by setting $\Sigma = r + \mathcal{O}(r^{-1})$, we find that, for $\lambda=0$,
\begin{align}
\begin{split}
\label{E:chargedsol}
	2 A &=  r^2 \left( 1 - \frac{r_h^4}{r^4} \left(1+\frac{1}{2} \frac{B^2}{r_h^4} \ln \left(\frac{r}{r_h}\right)\right) + \sum_{i=1}^3\frac{{q^i}^{2}}{3 r^6} \right)\,, \\
	\Sigma & = r\,, \qquad
	a^i = \mu^i - \frac{q^i}{r^2}\,,
\end{split}
\end{align}
is the most general solution to \eqref{E:EOM}. The horizon is located at $A(r_0)=0$ where we may choose $q^i = \mu^i r_0^2$ in order that the gauge fields are not singular at the bifurcation point. We will refer to this solution as the charged, magnetic black brane. When $\lambda \neq 0$ we must set $\mu_i = q_i = 0$ in order for \eqref{E:chargedsol} to solve the equations of motion. In this case the solution reproduces the one found in \cite{Donos:2011qt} in the context of gauged supergravity. We refer to the latter as an uncharged magnetic black brane. Note that in the absence of charges, $r_0=r_h$. 
 
The Hawking temperature for the charged solution is given by
\begin{equation}
\label{E:temperature}
	T =  \frac{r_0}{8\pi}\left(8-\frac{4}{3}\sum_{i=1}^3 \frac{{q^i}^2}{r_0^6} - \frac{B^2}{r_0^4} \right)\,,
\end{equation}
and the Bekenstein entropy per unit volume is
\begin{equation}
\label{E:gots}
	s = \frac{r_0^3}{4 G_N}\,.
\end{equation}
Using the AdS/CFT dictionary \cite{Sahoo:2010sp}, the stress tensor and charge currents associated with these solutions are given by 
{\tiny
\begin{align}
\begin{split}
\label{E:JTEV}
	16 \pi G_N \langle T^{mn} \rangle&= \begin{pmatrix}
		3 \left(r_h^4 + \frac{B^2}{4} \left(1-2 \ln\left(\frac{r_h}{r_c}\right)  \right)\right)& 0 & 0 & 0 \\
		0 & r_h^4 + \frac{B^2}{4}  \left(-1-2 \ln\left(\frac{r_h}{r_c}\right)  \right) & 0 & 0 \\ 
		0 & 0 & r_h^4 + \frac{B^2}{4} \left(-1-2 \ln\left(\frac{r_h}{r_c}\right)  \right)   & 0 \\
		0 & 0 & 0 & r_h^4 + \frac{B^2}{4}  \left(-1-2 \ln\left(\frac{r_h}{r_c}\right)  \right)    \\
	\end{pmatrix} \\ 
	8 \pi G_N \langle J_i^{m} \rangle&= \begin{pmatrix} q^i & 0 & 0 & 0 \end{pmatrix}\,
\end{split}
\end{align}}
with $r_c$ a renormalization scheme dependent parameter (see appendix \ref{A:prescription}).
Here, the roman indices $m,n=0,\ldots,3$ are spacetime coordinates of the boundary theory.
In the dual theory $\mu^i$ is the chemical potential associated with the $i$'th current and $B$ is an external magnetic field. Indeed, once the magnetic field doesn't vanish we find,
\begin{equation}
	T^{m}{}_{m} = - \frac{3 B^2}{32 \pi G_N}
\end{equation}
as expected \cite{Sahoo:2010sp}.

The thermodynamic constitutive relations for the stress tensor and current in the presence of a magnetic field have been computed in \cite{Huang:2011dc,Finazzo:2016mhm,Grozdanov:2016tdf,Hernandez:2017mch}. It is straightforward to generalize the former expressions to configurations with the $SO(3)_B$ spatio-magnetic symmetry under discussion. In the notation of \cite{Hernandez:2017mch} we find that in equilibrium
\begin{align}
\begin{split}
\label{E:Cmagnetic}
	T^{mn} &= \epsilon u^{m} u^{n} + \Pi \Delta^{mn} + \sum_i \alpha_i \left(B_i^{m} B_i^{n} - \frac{1}{3} \Delta^{mn} B_i^2\right) \\
	J_i^{m} & = \rho_i u^{m}
\end{split}
\end{align}
where $\Delta^{mn} = \eta^{mn} + u^{m}u^{n}$ is the projection orthogonal to the velocity field and $B_i^{m} = \frac{1}{2} \epsilon^{mnrs}u_{n} F^i_{rs}$ are the magnetic fields as seen in a local rest frame. The Gibbs-Duhem relation is given by
\begin{equation}
\label{E:GD}
	\epsilon+\Pi = s T + \sum_i \left(\mu_i \rho_i - \frac{2}{3} \alpha_i B_i^2\right)\,.
\end{equation}
For the case at hand we have $\alpha_i = \alpha$ for all $i$ from symmetry and $B_i^{m} = \delta_i^{m}B$ so that the last term proportional to $\alpha$ on the right of \eqref{E:Cmagnetic} vanishes. Comparing \eqref{E:Cmagnetic} and \eqref{E:GD} to \eqref{E:JTEV} and using \eqref{E:gots} and $u^m\partial_m = \partial_t$ we find
\begin{align}
\begin{split}
	16\pi G_N \, \epsilon &= 3 \left(r_h^4 + \frac{B^2}{4} \left(1-2 \ln\left(\frac{r_h}{r_c}\right)  \right)\right)  \\
	16 \pi G_N \, \Pi &= r_h^4 + \frac{B^2}{4} \left(-1-2 \ln\left(\frac{r_h}{r_c}\right)  \right) \\
	32\pi G_N \alpha & = -1 + 2 {\ln \left(\frac{r_0}{r_c}\right)}  \\
	8 \pi G_N \rho_i & = {\mu_i r_h^2} \,.
\end{split}
\end{align}

\section{Quasi normal modes of uncharged black branes}
\label{S:QNM}
The uncharged magnetic black brane solutions given by \eqref{E:chargedsol} with $q_i=0$ possess a resdiual $SO(3)_B$ symmetry (the diagonal of $SO(3)_f$ and $SO(3)$ rotational symmetry). When studying perturbations of these solutions it is convenient to decompose the perturbations into representations of this $SO(3)_d$ symmetry. We denote the metric perturbations by $\delta g_{\mu\nu} = r^2 e^{-i \omega t} \gamma_{\mu\nu}$ and the gauge field perturbations by $\delta A_{i\,\mu} = e^{-i \omega t} a_{i\,\mu}$. We will work in a gauge where $\gamma_{\mu r}=0$  and $a_{i\,r}=0$. In what follows we will use roman indices $i,j = 1,2,3$ to denote both spatial indices and $SO(3)_f$ indices. Thus, for example, $a_{i\,j}$ denotes the $j$'th spatial component of the $i$'th flavor type of the gauge field fluctuation.

Metric perturbations can be decomposed into symmetric traceless modes of the form $\gamma_{T\,ij} = g_{ij} - \frac{1}{3} g_{kj}\delta^{kl} \delta_{ij}$, vector perturbations of the form $\gamma_{V\,i} = \gamma_{0i} r_h$, and scalar perturbations, $\gamma_{S\,1} =  \gamma_{00}$ and $\gamma_{S\,2} = \gamma_{jk}\delta^{jk}$. We decompose gauge field perturbations in a similar way: symmetric traceless modes $a_{T\,ij} = a_{ij}+a_{ji} - \frac{2}{3} a_{kl}\delta_{kl} \delta_{ij}$, vector and pseudovector modes $a_{V\,i} = a_{i\,0}$ and $\tilde{a}_{V\,i} = \epsilon_{i}{}^{jk}a_{jk}$ (respectively), and a scalar mode $a_S = \delta^{ij}a_{ij}$.

Since the Chern-Simons term (which breaks the residual $SO(3)_B$ symmetry) does not affect the Einstein-Maxwell equations we find that the symmetric traceless modes of the metric, $\gamma_T(\rho)$, decouple and satisfy
\begin{subequations}
\label{E:Perturbationequations}
\begin{multline}
\label{E:MetricTensor}
	\rho ^2 \left(\beta ^2 \ln \rho-2 \rho ^4+2\right) \gamma _T'' +\rho  \left(\beta ^2 \ln \rho+\beta ^2-10 \rho ^4+4 i \rho ^3 \Omega +2\right) \gamma _T' 
	\\
	+ \left(2\beta ^2+6 i \rho ^3 \Omega \right) \gamma _T = 0
\end{multline}
where we have defined
\begin{equation}
\label{E:normalized}
	\rho = r/r_h\,, \qquad
	B = \beta r_h^2\,, \qquad
	\omega = \Omega r_h\,.
\end{equation} 
On the other hand, the tensor modes satisfy
\begin{multline}
\label{E:GaugeTensor1}
	\rho^2 \left(\beta ^2 \ln \rho -2 \rho ^4+2\right) a_{T\,ij}''
	-\rho  \left(\beta ^2 \ln \rho -\beta ^2+6 \rho ^4-4 i \rho ^3 \Omega +2\right)a_{T\,ij}' 
	\\
	+  2 i \rho ^3 \Omega  a_{T\,ij}
	=\begin{cases}
		0 & i = j \\
		-64 \beta ^2 \lambda ^2 a_{Tij} & i \neq j
	\end{cases}\,.
\end{multline}
The expression on the right hand side of this equation is due to the presence of the Chern-Simons term which breaks the $SO(3)_B$ symmetry; it vanishes for $\lambda=0$.

For the vector sector we find
\begin{equation}
	\rho^3 a'_{V\,i} = -4 \lambda \beta a_{T\,jk} \qquad \hbox{for $i\neq j \neq k$}\,.
\end{equation}
The non tensorial nature of this equation is, as before, a result of the Chern-Simons term.
The remaining vector equations are
\begin{multline}
	-i \rho  \left(\beta ^2 \ln \rho -2 \rho ^4+2\right) \tilde{a}_V'' 
	+ i  \left(\beta ^2 \ln \rho -\beta ^2+6 \rho ^4-4 i \rho ^3 \Omega +2\right)\tilde{a}_V'
	+2 \rho ^2 \Omega  \tilde{a}_V
	\\
	=
	-4 i \beta  \rho ^3 \gamma _V'(\rho )
	-4 i \beta  \rho^2 \gamma _V(\rho )\,, \label{E:GaugeVector1}
\end{multline}
and
\begin{equation}
	-{\beta   \left(\beta ^2 \log (\rho )-2 \rho^4+2\right)}\tilde{\alpha}_V'
	-2 i \beta \rho^2  \Omega  \tilde{\alpha}_V
	=
	-2 i \rho ^6 \Omega  \gamma_V'
	-4 \beta ^2 \rho^2 \gamma_V \,. \label{E:GaugeVector2}
\end{equation}

Finally, for the scalar sector we have
\begin{align}
\begin{split}
	2 \rho  \left(\beta ^2 \log (\rho )-2 \rho ^4+2\right) \gamma _{S\,2}'
	+\left(4 \beta ^2 \log (\rho )-\beta^2+4 i \rho ^3 \Omega +8\right)\gamma _{S\,2} 
	&=
	-4 \rho ^4 \gamma _{S\,1} 
	\\
	8 \rho ^6 \left(-\beta ^2 \ln\rho+2 \rho^4-2\right) \gamma_{S\,1}''
	+56 \rho ^5 \left(-\beta ^2 \ln\rho+2 \rho ^4-2\right) \gamma_{S\,1}'
	+E_1 \gamma_{S\,1} 
	&=E_2 \gamma_{S\,2} 
\end{split}
\end{align}
with
\begin{align}
\begin{split}
	E_{1} &=  4 \rho ^4 \left(-22 \beta ^2 \ln\rho+3 \beta ^2+20 \rho ^4+12 i \rho ^3 \Omega -44\right) \\
	E_2 &= 24 \beta ^4 \ln^2\rho +3 \beta ^4-\beta ^2 \left(68-20 \rho^4\right)-\left(34 \beta ^4-48 \beta ^2 \left(\rho ^4+2\right)\right) \ln\rho+48 \left(\rho ^6 \Omega ^2+2 \rho ^4+2\right)\,.
\end{split}
\end{align}
and
\begin{equation}
	\rho^2 \left(\beta ^2 \ln \rho -2 \rho ^4+2\right) a_{S}''
	-\rho  \left(\beta ^2 \ln \rho -\beta ^2+6 \rho ^4-4 i \rho ^3 \Omega +2\right)a_{S}' 
	\\
	+  2 i \rho ^3 \Omega  a_{S}
	=0
\end{equation}
\end{subequations}

In order to find the quasi normal modes associated with the magnetic black hole we must solve these equations with boundary conditions such that the modes are ingoing at the horizon and are not sourced on the boundary. 

The equations for the metric and gauge field perturbations \eqref{E:Perturbationequations} together with the ingoing boundary conditions reduce to Sturm-Liouville-type equations and can be satisfied only for particular values of the frequency $\omega$. For generic values of $\beta$ and $\lambda$ one must resort to numerics in order to compute the quasi normal frequencies. However, for particular values of $\beta$ and $\lambda$ some analytic solutions are available.

\subsection{The limit of zero magnetic field} \label{S:smallMagenticField} 
Consider the $\beta \to 0$ (zero magnetic field) limit of equations \eqref{E:Perturbationequations}. In this case the quasi normal modes reduce to those of the Schwarzschild AdS black brane which have been studied in \cite{Kovtun:2005ev}. One finds that the modes for the gauge field decouple and can be solved for analytically. 
In particular, one finds that $a_0(\rho)$, (with $a$ denoting either $a_T$, $\tilde{a}_V$ or $a_S$ at $\beta=0$) satisfies
\begin{equation}
\label{E:SLeqn}
	 (p a_0')' + q a_0 = 0
\end{equation}
where
\begin{align}
\begin{split}
\label{E:pq}
	p(\rho) &= -e^{-i \Omega_0 \arctan(\rho)}(\rho^4-1)\left(\frac{2}{1+\rho}-1\right)^{-\frac{i \Omega_0}{2}}\rho^{-1}
	\\
	q(\rho) &= ie^{-i \Omega_0 \arctan(\rho)}\left(\frac{2}{1+\rho}-1\right)^{-\frac{i \Omega_0}{2}}\Omega_0
\end{split}
\end{align}
and $\Omega_0 = \lim_{\beta \to 0}\Omega$.
The solution to \eqref{E:SLeqn} with ingoing boundary conditions is given by
\begin{align}
\begin{split}
\label{E:B0modes}
	a_0(\rho) &= C_0 (1- i \rho)^{-n(1-i)}(1+\rho)^{-n(1+i)}{}_2 F_1\left(1-n,-n,1-n(1+i);\frac{1}{2}(1-\rho^2)\right)\,,
	\\
	\Omega_0 &= 2n(1-i)
\end{split}
\end{align}
where $n \geq 1$ is an integer and $C_0$ a constant. There is an additional family of solutions with $\Omega = 2n(-1-i)$ and an associated $a(\rho)$ given by the conjugate of the one in \eqref{E:B0modes}.

One can now work perturbatively in $\beta$ to obtain corrections to the quasinormal frequency $\Omega$. In particular, let us write 
\begin{equation}
\label{E:expansion}
	a = a_0 + \beta^2 a_2 + \mathcal{O}(\beta^4)
	\qquad
	\Omega= \Omega_0 + \beta^2 \Omega_2 + \mathcal{O}(\beta^4)\,.
\end{equation}
Inserting \eqref{E:expansion} into \eqref{E:Perturbationequations} one finds
\begin{equation}
\label{E:VeqnB0O2}
	(p a_2')' + q a_2 = S 
\end{equation}
where
\begin{equation}
	S = \frac{1}{2\rho^3} 
		e^{-i\Omega_0 \arctan(\rho )}
		\left(\frac{2}{\rho +1}-1\right)^{-\frac{i \Omega_0}{2}}  \sigma
\end{equation}
with
{\tiny
\begin{align}
\begin{split}
	\sigma = \begin{cases} 
		 -\left( 64 \lambda ^2 + 2 i \rho^3 {\Omega_2} + \frac{i \rho ^3 {\Omega_0} \ln  \rho}{\rho^4-1}\right) a_0 
		 -\rho \left(4 i  \rho ^3 \Omega_2 + 1 -\frac{2 \rho ^3 (2 \rho -i \Omega_0) \ln \rho}{\rho ^4-1}\right)a_0' 
    & \hbox{$T$ modes $i\neq j$} \\
		 \left(4-2i\rho^3 \Omega_2 - \frac{i \rho^3 \Omega_0 \ln\rho}{\rho^4-1}\right) a_0 +\left(\frac{-4 i+4 i\rho^4 - \rho^3 \Omega_0 - 4 i \rho^6\Omega_0 \Omega_2}{\rho^2 \Omega_0} + \frac{2 \rho^4(2\rho-i \Omega_0)\ln\rho}{\rho^4-1} \right) a_0' +4\rho^3 \gamma_1 
	& \hbox{$\tilde{V}$ modes} \\
		 -\left( 2 i \rho^3 {\Omega_2} + \frac{i \rho ^3 {\Omega_0} \ln  \rho}{\rho^4-1}\right) a_0 
		 -\rho \left(4 i  \rho ^3 \Omega_2 + 1 -\frac{2 \rho ^3 (2 \rho -i \Omega_0) \ln \rho}{\rho ^4-1}\right)a_0'  & \hbox{$S$ modes and $T$ modes with $i=j$.}	
	\end{cases}\,.
\end{split}
\end{align}
}

and $\gamma_1$ is the leading component of an expansion of $\gamma_V$ in $\beta$, viz., $\gamma_V = \beta \gamma_1 + \mathcal{O}(\beta^3)$. The expression for the source term of the pseudo-vector components is supplemented by
\begin{equation}
	\rho^6 \Omega_0 \gamma_1' = \rho^2 \Omega_0 a_0 + i (\rho^4-1)a_0' \,.
\end{equation}
Since $a_0$ satisfies the boundary conditions for $a$, we can look for a solution where
\begin{equation}
\label{E:a2bc}
	a_2(\infty)=0 \qquad a_2(1)=0\,.
\end{equation}
(Note that we could also set $a_2(1)$ to take on any finite non zero value by adding to $a_2$ the solution to the homogenous equation \eqref{E:B0modes}.)

To solve \eqref{E:VeqnB0O2} we compute the relevant Green's function. The solutions to the homogenous equation 
$
	(p h')' + q h = 0
$
with $p$ and $q$ as in \eqref{E:pq} and $\Omega_0 = 2n(1-i)$ are
\begin{align}
\begin{split}
	h_1 &= (1- i \rho)^{-n(1-i)}(1+\rho)^{-n(1+i)}{}_2 F_1\left(1-n,-n,1-n(1+i);\frac{1}{2}(1-\rho^2)\right)\\
	h_2 & = (1- i \rho)^{-n(1-i)}(-1+\rho)^{n(1+i)}{}_2 F_1\left(1+i n,i n,1+n(1+i);\frac{1}{2}(1-\rho^2)\right)\,.
\end{split}
\end{align}
These solutions satisfy
\begin{equation}
	h_1  \xrightarrow[\rho\to1]{} \mathcal{O}((\rho-1) ^0)  , 
	\qquad
	h_2 \xrightarrow[\rho\to1]{} \mathcal{O}\left( (\rho-1)^{(1+i)n} \right)\,,
\end{equation}
near the horizon and
\begin{equation}
	h_1 \xrightarrow[\rho\to\infty]{} \mathcal{O}(\rho^{-2}) \,,
	\qquad
	h_2  \xrightarrow[\rho\to\infty]{}  \mathcal{O}(\rho^0) \,,
\end{equation}
near the asymptotic boundary. Thus, the most general solution to \eqref{E:VeqnB0O2} is
given by
\begin{equation}
\label{E:gota2}
	a_2(\rho) = A h_1(\rho) + B h_2(\rho) + {C} h_1(\rho) \int_1^{\rho} h_2(x)S(x)dx + {C}h_2(\rho) \int_{\rho}^{\infty} h_1(x) S(x) dx\,.
\end{equation}
where ${C}$ is a constant which can be read off the Wronskian of $h_1$ and $h_2$ and $A$ and $B$ are integration constants. 

For the tensor modes we find that
\begin{equation}
	h_1(x) S(x) \xrightarrow[x\to 1]{}  \mathcal{O}\left((x-1)^{-(1+i)n}\right)\,,
	\qquad
	h_2(x) S(x) \xrightarrow[x\to 1]{}  \mathcal{O}\left((x-1)^0\right)\,,
\end{equation}	
and
\begin{equation}
	h_1(x) S(x) \xrightarrow[x \to\infty]{} \mathcal{O}\left(x^{-4}\right)\,,
	\qquad
	h_2(x) S(x) \xrightarrow[x\to\infty]{}  \mathcal{O}\left(x^{-2}\right)\,.
\end{equation}
The first integral on the right hand side of \eqref{E:gota2} is well defined and converges in the $\rho \to \infty$ limit while the second integral on the right hand side of \eqref{E:gota2} vanishes as we near the boundary. Thus, 
\begin{equation}
	\lim_{\rho \to \infty}a_2(\rho)  = B h_2(\infty)
\end{equation}
suggesting that we must set $B=0$. 

Near the horizon the first integral on the right hand side of \eqref{E:gota2} vanishes but the second integral is divergent. To properly evaluate $a_2(\rho)$ in the $\rho \to 1$ limit we write
\begin{equation}
\label{E:divergenth1S1}
	h_1(x) S(x) = (1-x)^{-n i} \sum_{k=1}^{n} d_k x^{-2}(1-x)^{-k} + \Sigma(x)
\end{equation}
where $\Sigma(x) \xrightarrow[x \to 1]{} \mathcal{O}((1-x)^{-ni+0})$. That is, the sum on the right of \eqref{E:divergenth1S1} leads to a divergent integral when evaluated close to $x=1$. We can now regulate the last expression on the right hand side of \eqref{E:gota2} by writing
\begin{multline}
\label{E:regulatedh1S1}
	h_2(\rho) \int_{\rho}^{\infty} h_1(x) S(x) dx = h_2(\rho) \int_{\rho}^{\infty} \Sigma(x) dx  
	+ h_2(\rho)\sum_{k=1}^{n} d_k  \int_{\rho}^{\infty} x^{-2}(1-x)^{-n i} (1-x)^{-k} dx\,.
\end{multline}
We find
\begin{multline}
\label{E:otherconstraint}
	\lim_{\rho \to 1} h_2(\rho) \int_{\rho}^{\infty} h_1(x) S(x) dx \\
	= \left( \int_{1}^{\infty} \Sigma(x) dx + \sum_{k=1}^{n} d_k  \pi  (n-i k) (\coth (\pi  n)+1) \right) \lim_{\rho \to 1} h_2(\rho) +\left(\hbox{constant}\right) \,.
\end{multline}
To avoid outgoing modes at the horizon we must set the term in parenthesis on the right had side of \eqref{E:otherconstraint} to zero. This gives a constraint of the form
\begin{equation}
\label{E:gotO2}
	N_n + L_n \lambda^2 + W_n \Omega_2 = 0
\end{equation}
where, unfortunately, we have not managed to obtain an analytic expression for $N_n$, $L_n$ and $W_n$. Numerical estimates for these parameters can be found in table \ref{T:estimates}. Solving \eqref{E:gotO2} for $\Omega_2$ gives us the subleading correction (in $B$) to the tensor quasi normal modes.
\begin{table}[hbt]
\begin{center}
\begin{tabular}{| l | l | l | l |}
\hline
	$n$ & $N_n$ & $L_n$ & $W_n$ \\
\hline
	$1$ & $1.69018-0.641891 i$ & $-100.719-14.5216 i $ & $5.78517-5,78517 i $ \\
	$2$ & $3.91912+2.82223 i$ & $-67.6231-100.08 i$ & $9.3711+1.33873 i$ \\
	$3$ & $0.274092+6.71841 i$ & $29.6049-104.382 i$ & $4.30775+7.24068 i$ \\
	$4$ & $-5.47406+4.8077 i$ & $79.2794-35.2899i$ & $-2.28829+6.30677 i$ \\
\hline
\end{tabular}
\end{center}
\caption{\label{T:estimates} Numerical evaluation of the expressions in \eqref{E:gotO2} which determine subleading corrections to the quasinormal frequencies of the tensor modes in \eqref{E:B0modes}}.
\end{table}

The tensor quasi normal modes are the most interesting for our current purposes since they are sensitive to the Chern-Simons term we have introduced. From the point of view of the dual boundary theory they get modified due to an 't Hooft anomaly. The vector and scalar quasi normal modes may also be computed by similar means.

\subsection{Large Chern-Simons coupling}
Another interesting limit one can take is $\lambda\to\infty$ while keeping $B \lambda = M^2$ fixed (see \cite{Haack:2018ztx}). This corresponds to taking the $\lambda \to \infty$ limit of \eqref{E:ansatz} but keeping $A_{\mu}\lambda$ fixed. In this case, the Maxwell and Chern-Simons terms become subleading relative to the Einstein-Hilbert term, leading to a probe limit where the gauge field explores a background AdS black brane geometry. If we further take the zero temperature limit, $\beta=\sqrt{8}$ ($r_h  = \frac{M}{\sqrt{\lambda\sqrt{8}}}$) we find that the symmetric tensor field fluctuations satisfy 
\begin{equation}
\label{E:SLteqn}
	 \left( \tilde{p}  \tilde{a}'_{1} \right)'  + \tilde{q} \tilde{a}_{1} = 0 
\end{equation}
where
\begin{equation}
	\tilde{p} =  e^{\frac{2 i  \tilde{\Omega}_0}{\tilde{\rho}}} {\tilde{\rho}^3}
	\qquad
	\tilde{q} = - e^{\frac{2 i  \tilde{\Omega}_0}{\tilde{\rho}}} \left(i \tilde{\Omega}_0 + \frac{1}{\tilde{\rho}^3}\right)
\end{equation}
and we have defined
\begin{equation}
	\tilde{\rho} = \frac{\rho}{4\sqrt{\lambda}} = \frac{r}{32^{\frac{1}{4}}M}
	\qquad
	\tilde{\Omega} = \frac{\Omega}{4\sqrt{\lambda}} = \frac{\omega}{32^{\frac{1}{4}}M}
\end{equation}
with $\tilde{a}_1$ denoting $a_{T\,ij} \lambda$ with $i \neq j$ in the $\lambda \to \infty$ limit, and $\tilde{\Omega}_0 = \lim_{\lambda \to \infty} \tilde{\Omega}$. Solving \eqref{E:SLteqn} with the boundary conditions $\tilde{a}_1(\infty) = 0$ and $\lim_{\lambda \to \infty} \tilde{a}_1\left(\frac{1}{4\sqrt{\lambda}}\right) = \hbox{finite}$ we find
\begin{equation}
\label{E:leadingtildeO}
	\tilde{a}_{1}(\tilde{\rho}) = \tilde{C}_0 e^{-\frac{1}{2 \tilde{\rho}^2}  -i \frac{\tilde{\Omega}_0 }{\tilde{\rho} }}L_{n}^{-1}\left(\tilde{\rho}^{-2}\right)
	\qquad
	\tilde{\Omega}_0 = \pm 2 \sqrt{n}\,.
\end{equation}
where $L_m^{\nu}(x)$ is a generalized Laguerre polynomial.

Working perturbatively in $\lambda$ we expand
\begin{equation}
	\tilde{a} = \frac{\tilde{a}_1}{\lambda} + \frac{\tilde{a}_{3} + \tilde{a}_{L\,3} \ln \lambda }{\lambda^{3}} + \ldots
\end{equation}
where $\tilde{a}$ represents $a_{T\,ij}$ with $i \neq j$,
\begin{equation}
	\tilde{\Omega} = \tilde{\Omega}_0 + \frac{\tilde{\Omega}_{2}+\tilde{\Omega}_{L,2} \ln \lambda}{\lambda^2} + \ldots \,,
	\qquad
	\beta = \sqrt{8}   - \frac{\pi  32^{\frac{1}{4}}}{M} \frac{{T_1} +T_{L\,1} \ln \lambda}{\lambda^2} + \ldots \,,
\end{equation}
so that $r_h(\lambda) = \frac{M}{\sqrt{\lambda}}\left(\frac{1}{8^{\frac{1}{4}}}+ \frac{\pi }{4 M }\frac{{T_1} +T_{L\,1} \ln \lambda}{\lambda^2} +\ldots \right)$ or $T = \frac{T_1+T_{L\,1}\ln\lambda}{\lambda^{\frac{5}{2}}} + \ldots$ with $T$ the temperature. As we will see shortly, the powers of $\lambda$ in the perturbative expansion are dictated by the structure of the perturbative equations of motion.

The resulting equations of motion for $a_2$ are
\begin{equation}
\label{E:a32eq}
	\left( \tilde{p} \tilde{a}'_{3/2} \right)'  + \tilde{q} \tilde{a}_{3/2} = \tilde{S}
\end{equation}
with
\begin{equation}
	\tilde{S} = e^{-\frac{2 i \tilde{\Omega}_0}{\tilde{\rho}}} \left( \left( i \tilde{\Omega}_{1/2} - \frac{8^\frac{1}{4} \pi T_1}{M \tilde{\rho}^3}  \right) \tilde{a}_1 - 2 i \tilde{\Omega}_{1/2} \tilde{\rho} \tilde{a}_1'\right)\,.
\end{equation}
The ingoing boundary conditions now translate to	
\begin{equation}
	\tilde{a}_{3}(\infty) = 0\,,
	\qquad
	\tilde{a}_{L\,3}(\infty) = 0\,,
	\qquad
	\tilde{a}_{3}\left(0 \right) = 0\,,
	\qquad
	\tilde{a}_{L\,3}\left(0 \right) = 0\,.
\end{equation}
where the last two equalities follow from the asymptotic behavior of $\tilde{a}_1$ at large $\tilde{\rho}$.

As in the previous limit, we compute the subleading corrections to the quasi normal modes by constructing the Greens function for the solution. The two linear solutions to \eqref{E:SLteqn} are given by
\begin{align}
\begin{split}
	\tilde{h}_1 &= e^{-\frac{1}{2 \tilde{\rho}^2}  -i \frac{2 \sqrt{n} }{\tilde{\rho} }}L_{n}^{-1}\left(\tilde{\rho}^{-2}\right) \\
	\tilde{h}_2 & =e^{-\frac{1}{2 \tilde{\rho}^2}  -i \frac{2 \sqrt{n} }{\tilde{\rho} }} n \left(
		e^{\frac{1}{\tilde{\rho}^{2}}} Q_n(\tilde{\rho}^{-2}) + Ei\left(\tilde{\rho}^{-2} \right){L_{n}^{-1}(\tilde{\rho}^{-2})} 
		\right)
\end{split}
\end{align}
where $Ei(x)$ is the exponential integral function $Ei(x)=-\int_{-x}^{\infty}\frac{e^{-z}}{z}dz$ and 
\begin{equation}
	Q_n(x) = \frac{1}{n!} \sum_{k=0}^{n-1} c_k (-x)^{n-k-1}
\end{equation}
is a polynomial of its argument with coefficients $c_k$ satisfying the recursion relation
\begin{equation}
	c_k = \frac{(n-k) (-k+n+1)}{-k+2 n+1}c_{k-1} 
	+\frac{(2 (n-k)+1) \Gamma (n+1) \Gamma (n+2)}{(-k+2 n+1) \Gamma (k+1) \Gamma (-k+n+1) \Gamma (-k+n+2)}
\end{equation}
with $c_0=1$. Near the horizon ($\tilde{\rho} \to 0$) we have
\begin{equation}
	\tilde{h}_1 = \mathcal{O}\left(e^{-\frac{1}{2\tilde{\rho}^2}}\tilde{\rho}^{-2n}\right)
	\qquad
	\tilde{h}_2 = \mathcal{O}\left(e^{\frac{1}{2\tilde{\rho}^2}}\tilde{\rho}^{-2(n-1)}\right)
\end{equation}
and near the boundary ($\tilde{\rho} \to \infty$) we have
\begin{equation}
	\tilde{h}_1 = \mathcal{O}\left(\tilde{\rho}^{-2}\right)
	\qquad
	\tilde{h}_2 = \mathcal{O}\left(\tilde{\rho}^0\right)\,.
\end{equation}
The general solution to \eqref{E:a32eq} is
\begin{equation}
	a_{3/2} = \tilde{A} \tilde{h}_1 + \tilde{B} \tilde{h}_2 + \tilde{C} \tilde{h}_1 \int_0^{\tilde{\rho}} \tilde{h}_2(x) \tilde{S}(x) dx + \tilde{C} \tilde{h}_2 \int_{\tilde{\rho}}^{\infty} \tilde{h}_1(x) \tilde{S}(x)  dx\,.
\end{equation}
C.f., \eqref{E:gota2}.

A short computation reveals that
\begin{equation}
	\tilde{h}_2(x) \tilde{S}(x) \xrightarrow[x\to\infty]{} \mathcal{O}(x^{-2})
\end{equation}
implying that $\tilde{B}=0$. Further,
\begin{equation}
	\tilde{h}_1(x) \tilde{S}(x) \xrightarrow[x\to 0]{} \mathcal{O}\left(e^{-\frac{1}{x^2}}x^{-4n+7}\right)
\end{equation}
so the second integral converges near the horizon. In order for the boundary conditions to be satisfied we must set
\begin{equation}
\label{E:hardintegral}
	\int_{0}^{\infty} \tilde{h}_1 \tilde{S} dx = 0\,.
\end{equation}
It is straightforward though somewhat tedious to solve \eqref{E:hardintegral}. We find
\begin{align}
\begin{split}
\label{E:solvedtildeO}
	0&= N_n - \frac{1+7 n^2}{16}\ln 2 - \frac{1+7 n^2}{64}  \gamma - \frac{8^{\frac{1}{4}}\pi T_1}{M} - \frac{2}{\sqrt{n}} \tilde{\Omega}_2  \\
	0&=-\frac{1+7n^2}{64} - \frac{8^{\frac{1}{4}}\pi T_{L\,1}}{M } - \frac{2}{\sqrt{n}} \tilde{\Omega}_{L\,2} \,,
\end{split}
\end{align}
where
\begin{equation}
	N_n = \frac{1}{384}\left(145+ \frac{174}{n-2}+\frac{48}{n-1}+\frac{3}{n}+105 n + 26 n^2 + 6(1+7 n^2)H_{n-3}\right)
\end{equation}
for $n\geq 3$, $H_m = \sum_{k=1}^m \frac{1}{k}$ are harmonic numbers and $\gamma$ represents Euler's constant, $\gamma = 0.5772\ldots$. The expressions for $N_1$ and $N_2$ can be obtained from the appropriate limits of $N_n$, viz., 
\begin{equation}
	N_1 = \frac{23}{128}\,,
	\qquad
	N_2 = \frac{227}{256}\,.
\end{equation}
The subleading corrections to the quasinormal mode frequencies can be read off from \eqref{E:solvedtildeO}. Note that they are real to the order we are working in. 


\subsection{Full solution}

\begin{figure}[h]
\begin{center}
{\includegraphics{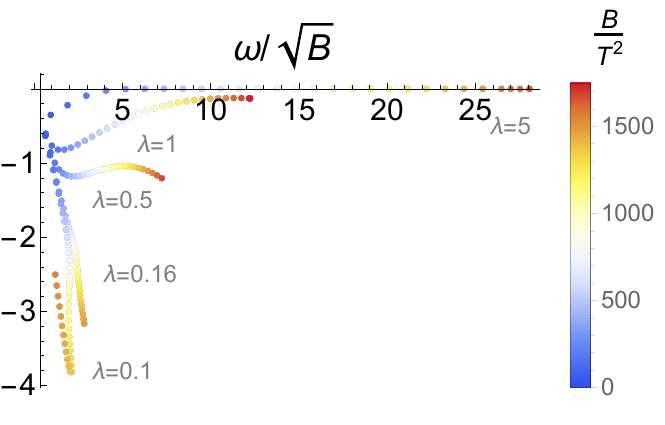}}
{\caption{ \label{F:plots} Complex values of the first quasi normal mode in units of the magnetic field, $B$, as a function of inverse temperature, $T$. As the temperature decreases the quasi normal modes with $\lambda \geq 1$ approach the real axis. Here $\lambda$ specifies the strength of the Chern-Simons term.}}
\end{center}
\end{figure}

Starting from the analytic solutions for small values of $\beta$ \eqref{E:B0modes}, we use a shooting algorithm to solve \eqref{E:GaugeTensor1} for increasingly larger values of $\beta$. Typical behavior of the quasi normal is exhibited in table \ref{F:plots}.

\begin{figure}[h]
\begin{center}
{\includegraphics{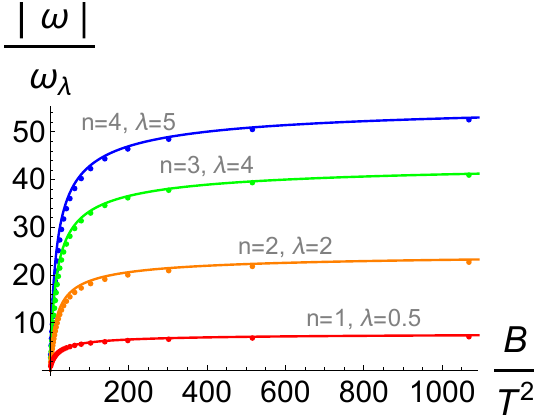}}
{\caption{ \label{F:largelfit}  A comparison of the absolute value of the quasi normal modes obtained numerically, $|\omega|$, denoted by circles, to the theoretical prediction obtained by taking the large $\lambda$ limit, $\omega_{\lambda}$, denoted by a solid line, c.f., \eqref{E:leadingtildeO} and \eqref{E:solvedtildeO}. Here $\lambda$ specifies the strength of the Chern-Simons coupling and $n$ the number of the quasi normal mode. Note that $\omega_{\lambda}$ is real while $\omega$ has a complex component.  }}
\end{center}
\end{figure}

When comparing the numerical value of the quasi normal modes to the theoretical, large $\lambda$ and small temperature (small $\sqrt{8} > \beta > 0 $) prediction (equations \eqref{E:leadingtildeO} and \eqref{E:solvedtildeO}), we find a surprisingly good match down to very low temperatures, and for values of $\lambda$ larger than, roughly, $1/2$. See figure \ref{F:largelfit}.

\begin{figure}[h]
\begin{center}
{\includegraphics{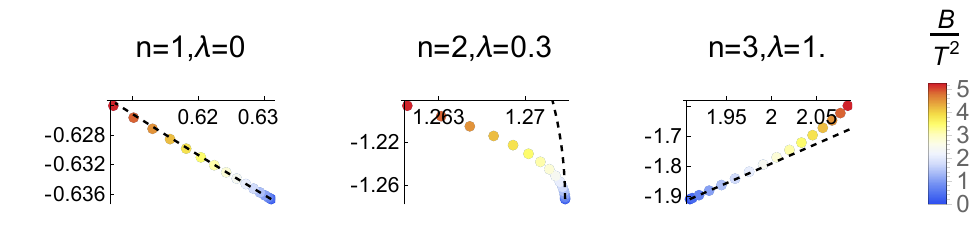}}
{\caption{ \label{F:smallfit}  A comparison of the value of the quasi normal mode to the small magnetic field approximation for various mode numbers and values of $\lambda$, the Chern-Simons term. The colored circles denote the location of the quasi normal modes, $\omega/T$, in the complex frequency plane as a function of magnetic field over temperature squared, $B/T^2$. The dashed line is the approximated value of the quasi normal mode following from \eqref{E:B0modes} and \eqref{E:gotO2}.  }}
\end{center}
\end{figure}

At small values of the magnetic field, we find that the approximation given in  \eqref{E:B0modes} and \eqref{E:gotO2} provides a good match to the data as long as the Chern-Simons coupling $\lambda$ is small. Once $\lambda$ is of order $1$ the small magnetic field approximation seems to break down rapidly. See figure \ref{F:smallfit}.



\section{Response to an electric field}
\label{S:Driving}

Our main goal in this work is to use the gauge gravity duality to understand how the boundary theory responds to an external electric field. Indeed, recall that a $U(1)$ gauge field in the bulk, $A^i$, is dual to a conserved current, $J^i$ in the boundary. These conserved currents are associated with global symmetries of the (boundary) theory, e.g., the $U(1)^3$ Cartan subgroup of the $SO(6)$ $R$ symmetry in $\mathcal{N}=4$ super Yang-Mills is dual to the three $U(1)$ currents in the STU ansatz of \cite{Behrndt:1998jd}.\footnote{While \eqref{E:chargedsol} with $\lambda=$ is a solution to type IIB supergravity, our analysis of the quasi normal modes may not be compatible with supergravity solutions---in that case extra scalar fields will be excited and their oscillations will contribute to the quasi normal modes. It is expected that the tensor and vector quasi normal modes which we are interested in won't be modified from the current ones.}

Leaning on the results of \cite{Sahoo:2010sp} (which rely, in turn on \cite{deHaro:2000vlm}), having an external electric field on the boundary implies that the boundary value of the associated gauge field $A^{\mu}$ should support a non trivial electric field. The current $J^{\mu}$ dual to $A^{\mu}$ in the presence of said electric field can then be read off the asymptotic behaviour of the $A^{\mu}$ as it approaches the boundary. Thus, to compute $J^{\mu}$ in the presence of an electric field we need to perturb the black hole solution of \eqref{E:chargedsol} such that the boundary value of the gauge field $A_{\mu}^i$ supports an electric field.

Before proceeding with the computation outlined above we would like to point out that the late time behavior of the current can be gleaned from the quasi normal mode analysis we have carried out in the previous section. Consider a charged black hole of the form \eqref{E:chargedsol} which is perturbed at some time $t=0$ by an external, time dependent, boundary electric field. Long after the black hole is perturbed we expect it to asymptote to its equilibrium solution \eqref{E:chargedsol} with deviations characterized by its quasi normal modes. Hence, if the temperature is small enough, and $\lambda$ large enough we expect to observe the long lived quasi normal modes of the black hole. From the perspective of the boundary theory we expect that after exciting the equilibrated system by a localized (in time) electric field, an oscillatory current will be observed even long after the lifetime of the excitation. Following \cite{Haack:2018ztx} we refer to this effect as anomalous resonance.


Demonstrating that the anomalous resonance effect takes place as suggested in the previous paragraph requires knowledge of the dynamics of perturbed black holes. Explicit solutions to the Einstein equations describing perturbed black holes are scarce and often rely on the existence of an underlying symmetry. Therefore, in order to observe the anomalous resonance effect we resort to numerics. The remainder of this section is divided in two. In \ref{SS:settingup} we provide some technical details regarding the numerical solution, and in \ref{SS:results} we present our results.

\subsection{Setting up the numerical problem}
\label{SS:settingup}
We will consider a setup where the black hole \eqref{E:chargedsol} is perturbed by non vanishing boundary electric fields $\vec{E}^1$ and $\vec{E}^2$ whose non vanishing components are given by $E = E^1_2 = E^2_1$ where, we remind the reader, the upper index is a flavor index and the bottom index is a spacetime index. This type of boundary electric field breaks the $SO(3)_B \times \mathbb{Z}_2$ symmetry of the black hole ansatz \eqref{E:ansatzsolution} (with $a^i=0$ and $\mathbb{Z}_2$ denoting parity) to a discrete $\mathbb{Z}_2$ subgroup involving rotations by ${\pi}/{2}$ in the $(x,y)$ plane and parity in the spatial coordinates. 

To implement the $\mathbb{Z}_2$ symmetry described above, we use an ansatz
\begin{subequations}
\label{E:Sansatz}
\begin{equation}
	ds^2 = -2 A(t,r) dt^2 + 2 d t d r + \Sigma^2(r,r) \left(g(t,r)dx^2 + g(t,r) dy^2 + g(t,r)^{-2} dz^2\right)
\end{equation}
for the line element and
\begin{equation}
	A^1 = a(t,r)dy+B y dz,\qquad
	A^2 = (a(t,r)+B z) dx\,\qquad
	A^3= \tilde{a}(t,r)dt+ B x d z\,,
\end{equation}
\end{subequations}
for the gauge fields. 

Using the boundary conditions
\begin{equation}
	\lim_{r\to\infty}2A = r^2,\,
	\qquad
	\lim_{r\to\infty} \Sigma = r,\,
	\qquad
	\lim_{r\to\infty}g = 1,\,
\end{equation}
ensures that the boundary metric is the Minkowski metric. Likewise,
\begin{equation}
	\lim_{{r} \to \infty} \epsilon^{ijk} \partial_j A_k^{\ell} = B \delta^{i\ell},\,
	\qquad
	 \lim_{{r} \to \infty} \partial_t a = -E(t) \,,
	 \qquad
	 \lim_{r\to\infty} \tilde{a} = 0\,.
\end{equation}
ensures a homogenous magnetic field $B$, and an electric field $E$ which breaks the $SO(3)\times \mathbb{Z}_2$ symmetry as described above.

With these boundary conditions, inserting the ansatz \eqref{E:Sansatz} into the equations of motion  \eqref{E:EOM} yields a near boundary expansion of the form
\begin{align}
\begin{split}
\label{E:BV}
	r_h^{-1} a & = a^{(0)}(t_h) + \frac{\partial_{t_h} a^{(0)}(t_h)}{\rho} - \frac{\partial_{t_h}^2 a^{(0)}}{2 \rho^2}\ln {\rho} + \frac{a^{(2)}(t_h) }{\rho^2} + \mathcal{O}(\rho^{-3}) \\
	r_h^{-1} \tilde{a} & = \frac{c+4 \beta  \lambda a^{(0)}(t_h)}{\rho^2} + \mathcal{O}(\rho^{-3})\\
	r_h^{-1} \Sigma &= \rho + \mathcal{O}(\rho^{-3}) \\
	g & = 1 - \frac{\partial_{t_h} a^{(0)}(t_h)}{12 \rho^4} \ln \rho + \frac{g^{(4)}}{\rho^4} + \mathcal{O}(\rho^{-5})\\
	r_h^{-2} A & = \frac{\rho^2}{2} - \frac{3 \beta^2 + 2 (\partial_{t_h} a^{(0)}(t_h))^2}{12 \rho^2} \ln \rho + \frac{A^{(2)}}{\rho^2} + \mathcal{O}(\rho^{-3}) \\
\end{split}
\end{align}
where we have defined the dimensionless quantities $\beta = B r_h^{-2}$, $\rho = r/r_h$ and $t_h = t r_h$ (recall that we are working in units where $\frac{r}{L^2}$ scales as energy)
and we have used the residual gauge freedom of our ansatz, $r \to r+\lambda(t)$, to set $\partial_r \Sigma = 1 + \mathcal{O}(r^{-1})$.

The undetermined integration parameters in \eqref{E:BV}, $a^{(0)}$, $a^{(2)}$, $c$, $g^{(4)}$ and $A^{(2)}$ determine the boundary electric field $E=E_2^1=E^1_2$, the associated expectation value of the (covariant) flavor currents, $J^{m}_{i\,cov}(t)$, and the  expectation value of the energy momentum tensor, $T^{mn}(t)$
\begin{align}
\begin{split}
\label{E:EVcurrents}
	E &= - r_h^2  a^{(0)\prime}(t_h) \\	
	8 \pi G_5 \langle J_{1\,cov}^m \rangle &= \delta^m_2 r_h^3 \left( a^{(2)} (t_h) - \frac{3}{8} a^{(0)''}(t_h) \right)\\
	8 \pi G_5 \langle J_{2\,cov}^m \rangle &= \delta^m_1 r_h^3 \left( a^{(2)} (t_h) - \frac{3}{8} a^{(0)''}(t_h) \right) \\
	8 \pi G_5 \langle J_{2\,cov}^m \rangle & = - \delta^m_0 r_h^3 \left(c+ 4 \beta \lambda a^{(0)}( t_h)\right)
\end{split}
\end{align}
and
\begin{align}
\begin{split}
\label{E:EVstress}
	16 \pi G_5 \langle T_{00} \rangle &= r_h^4\left( -6 A^{(2)}(t_h) 
		-\frac{1}{6}  \left( a^{(0)\prime}(t_h) \right)^2 
		+ \frac{3}{4} \beta^2  \right)\\
	16 \pi G_5 \langle T_{11} \rangle &  =r_h^4 \left( -2 A^{(2)}(t_h) 
		+ 4 g^{(4)}(t_h) 
		+ \frac{7}{36}\left( a^{(0)\prime}(t_h) \right)^2  
		-\frac{1}{4} \beta^2 \right) \\
	16 \pi G_5 \langle T_{33} \rangle &= r_h^4\left( -2 A^{(2)}(t_h) - 8 g^{(4)} (t_h)
		+ \frac{4}{9}  \left(  a^{(0)\prime}(t_h) \right)^2  
		-\frac{1}{4} \beta^2 \right)  \\
\end{split}
\end{align}
with the other components vanishing. (See appendix \ref{A:prescription}; in writing \eqref{E:EVcurrents} and \eqref{E:EVstress} we have also chosen a scheme where the logairthms in \eqref{E:TEV} vanish.)
Note that (as is always the case) energy momentum conservation and current conservation are compatible with the equations of motion, 
\begin{equation}
\label{E:A2dot}
	\partial_{t_h} A^{(2)} = \frac{2}{3}a^{(2)} \partial_{t_h} a^{(0)} - \frac{7}{18} \partial_{t_h} a_0 \partial_{t_h}^2 a^{(0)} \,.
\end{equation}

The values of $a^{(0)}$, $c$, $a^{(2)}$, $g^{(4)}$ and $A^{(2)}$ control the thermal expectation values of the currents and stress tensor. The coefficient $a^{(0)}$ is determined from the boundary condition that $E=-\partial_t a^{(0)}$. Likewise, requiring that the covariant current vanish before turning on the electric field implies $c=0$. The remaining functions can not be determined by a near boundary expansion and one must resort to numerics to determine their explicit values.

To numerically solve the resulting set of equations we use the methods of \cite{Chesler_2014} to rewrite the Einstein equations as a set of nested linear equations. Indeed, if we replace time derivatives with outgoing null derivatives,
\begin{equation}
\label{E:dplus}
	d_+ = \partial_t + A\partial_r\,,
\end{equation}
then, given values for $\Sigma(t_0,r)$, $g(t_0,r)$ $a(t_0,r)$ and $A^{(2)}(t_0)$ at some time $t_0$,
the equations for $d_+\Sigma(t_0,r)$, $d_+g(t_0,r)$, $d_+a(t_0,r)$, $A(t_0,r)$ and $\partial_t A^{(2)}(t_0)$ become a set of nested linear equations which we can easily solve sequentially. With the solution for the above variables available, we can step forward in time using \eqref{E:dplus}. Note that $\tilde{a}$ can be completely removed from the equations of motion using
\begin{equation}
	\partial_r \tilde{a} = - \frac{8 B \lambda a}{\Sigma^3}\,.
\end{equation}

In slightly more detail, we define the barred variables,
\begin{align}
\begin{split}
	r_h^{-1} \Sigma &= \rho +\rho^{-2} \overline{\Sigma}  \\
	r_h^{-2} A &= \frac{1}{2}\rho^2 - \frac{3 \beta + 2 \left(  a^{(0)\prime}\right)^2}{12} \rho^{-2}\ln \rho + \rho^{-2} \overline{A} \\
	r_h^{-2} d_+\Sigma &= \frac{1}{2} \rho^2 - \frac{3 B + 2 \left( a^{(0)\prime}\right)^2}{12} \rho^{-2} \ln \rho  + \rho^{-2} \overline{d_+\Sigma} \\
	r_h^{-1} d_+ g & = \frac{1}{6} \left( a^{(0)\prime\prime}\right)^2 \rho^{-3} \ln \rho - \frac{\partial_t a^{(0)\prime}  a^{(0)\prime\prime}}{6} \rho^{-4} \ln \rho + \rho^{-3} \overline{d_+g}\\
	r_h^{-2} d_+a & = \frac{ a^{(0)\prime\prime}}{2} \rho^{-1} \ln \rho + \frac{ a^{(0)\prime\prime\prime}}{4} \rho^{-2} \ln \rho + \frac{ a^{(0)\prime\prime\prime\prime}}{2} \rho^{-3} \ln \rho + \rho^{-1} \overline{d_+a}\,.
\end{split}
\end{align}
Our initial data involves values for $\overline{\Sigma}$, $\overline{g}$, $\overline{a}$ and $A^{(2)}$ at some time $t_0$. Given these inital values we can solve the first order linear ordinary differential equation for $\overline{d_+\Sigma}$ with the boundary conditions
\begin{equation}
	\overline{d_+\Sigma}\Big|_{\rho \to \infty} = \frac{1}{2} \left(3 \rho^2 \partial_\rho \overline{\Sigma}\Big|_{\rho\to\infty} + 2 A^{(2)}\right)\,.
\end{equation}
Next we solve for $d_+g$ and $d_+a$. These form a set of two coupled first order ordinary differential equations. The boundary conditions we use are
\begin{equation}
	\overline{d_+g}\Big|_{{\rho}\to\infty} = - 2 g^{(4)} - \frac{\left( a^{(0)\prime}\right)^2}{24}\,,
	\qquad
	\overline{d_+a}\Big|_{{\rho}\to\infty} = \frac{3}{4} a^{(0)\prime\prime} - \frac{1}{2} A^{(2)} \,.
\end{equation}
Next up is a second order differential equation for $\overline{A}$, as boundary conditions we use
\begin{equation}
	\overline{A}\Big|_{{\rho}\to\infty} = A^{(2)}
\end{equation}
and $ A^{(0)} =0$. Finally, we evolve $A^{(2)}$ using \eqref{E:A2dot}.

In practice at each time we used $21$ Gauss-Lobatto grid points in the radial direction which covered ${0.91} \leq {\rho} < \infty$ and expanded all functions in Chebyshev cardinal functions, supported on those grid points. We have checked that ${\rho}={0.91}$ is always on or inside the event horizon. We used a fourth order Runge-Kutta algorithm to integrate forward in time.

\subsection{Results}
\label{SS:results}

To demonstrate the existence of an anomalous resonance effect we excite the system at some initial time $t$ by turning on an electric field
\begin{equation}
\label{E:dtGaussian}
	E(t) = - \frac{E_0}{2} \frac{\partial}{\partial t} e^{-\frac{t^2}{2\tau^2}}\,.
\end{equation}
(We have used \eqref{E:dtGaussian} instead of a Gaussian to disentangle the anomalous resonance effect from another effect associated with the anomaly referred to as an ``anomalous trailing'' effect. See \cite{Haack:2018ztx}.) Going to Fourier space, this corresponds to an electric field peaked around a frequency of $\pm 1/\tau$. Thus, we expect to observe an anomalous resonance effect whenever $\hbox{Re}(\omega_n) \sim \frac{1}{\tau}$ with $\omega_n$ the $n$'th quasi normal mode, and $-\hbox{Im}(\omega_n) \ll 1$. In practice, we found that the 1st quasi normal mode is excited as long as $\hbox{Re}\left( \omega_1 \right) \lesssim \frac{4}{\tau}$.

\begin{figure}[h]
\begin{center}
{\includegraphics[scale=0.8]{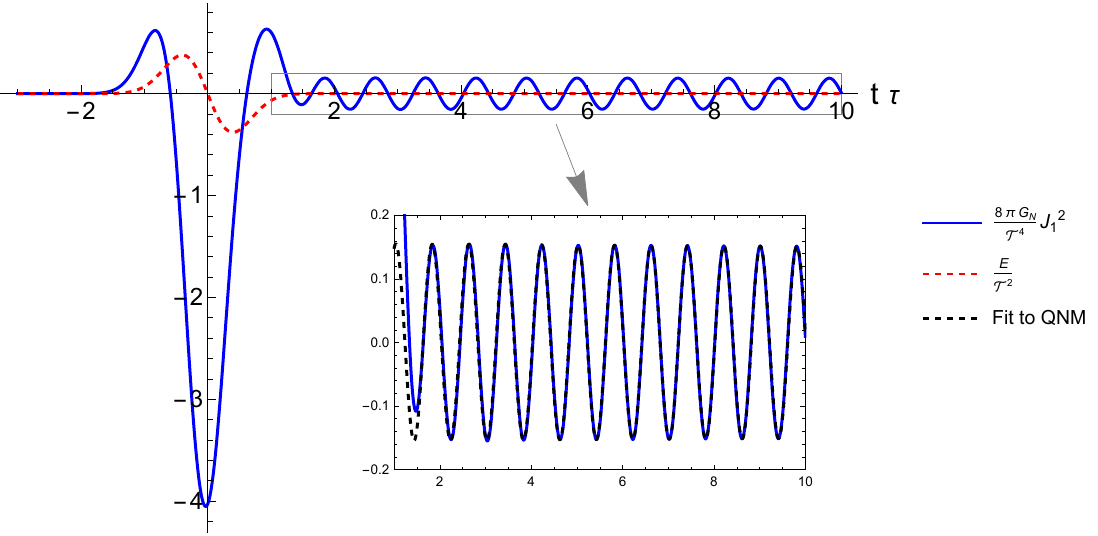}}
{\caption{ \label{F:currentar} The expectation value of the current as a function of time (blue) in the presence of an external electric field (dashed red) for $\lambda=3/2$, $B=2\mathcal{T}^2$, and an electric field described by \eqref{E:dtGaussian} with $E_0 = 1/2 \mathcal{T}^2$ and $\tau \mathcal{T}=2/5$. The Fourier transform of the external electric field is supported at $1/\tau \sim \hbox{Re}(\omega_1)/3$ with $\omega_1$ the quasi normal mode (QNM) associated with the gauge field. The excitation of the quasi normal mode manifests in the boundary theory as an oscillatory current.}}
\end{center}
\end{figure}

Our initial run involves an electric field of strength $E_0 = 0.5 \mathcal{T}^2$ and width $\tau \mathcal{T} = 2/5$, a magnetic field of $B = 2 \mathcal{T}^2$, and an anomaly with $\lambda = 3/2$ where we have introduced $\mathcal{T} = 2 \pi T_0$ with $T_0$ the initial temperature prior to the excitation. For these parameters we find that the final temperature of the system, long after the electric field has been turned on, asymptotes to $T_f \sim 1.01 T_0$. In this case the lowest quasi normal mode of the late time solution is given by $\omega_1 = 7.87-0.003 i$, so that $Re(\omega_1) \sim 3.1/ \tau$. This is sufficient to generate an anomalous resonance as exhibited in figure \ref{F:currentar}. 

\begin{figure}[h]
\begin{center}
{\includegraphics[scale=0.8]{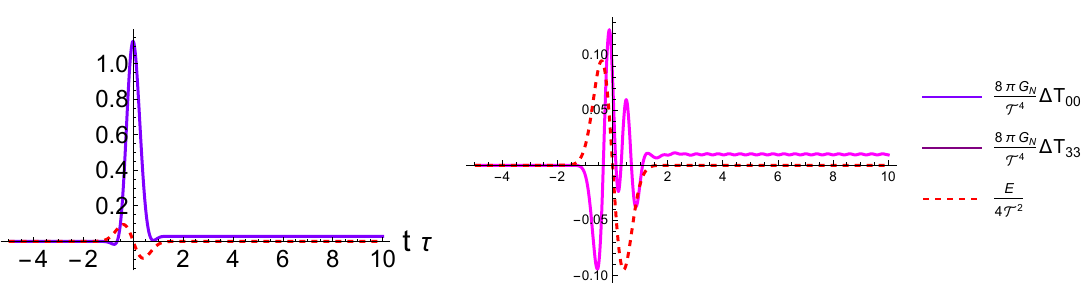}}
{\caption{ \label{F:metricar} We plot components of the stress tensor  (purple and pink) relative to the stress tensor for an unpertrubed magnetic black brane at $\lambda=3/2$, $B=2\mathcal{T}^2$, and an electric field described by \eqref{E:dtGaussian} with $E_0 = 1/2 \mathcal{T}^2$ and $\tau \mathcal{T}=2/5$ (dashed red). It seems that $T_{33}$ exhibits small oscillations at a frequency which is double that of the first quasi normal mode of the gauge field.}}
\end{center}
\end{figure}

Since the resulting late time temperature change is small, $(T_f-T_0)/T_0 \sim 0.01$, the backreaction of the geometry to the electric field is almost negligible outside the transient region where the electric field is non zero. 
Away from the transient region the spatial components of the metric fluctuate at a small amplitude, and at a frequency which is double that of the QNM of the gauge field, probably due to the non-linear nature of the Einstein equations. See figure \ref{F:metricar}. We have checked that these modes decay at a very low rate, as expected.

\begin{figure}[h]
\begin{center}
{\includegraphics[scale=0.8]{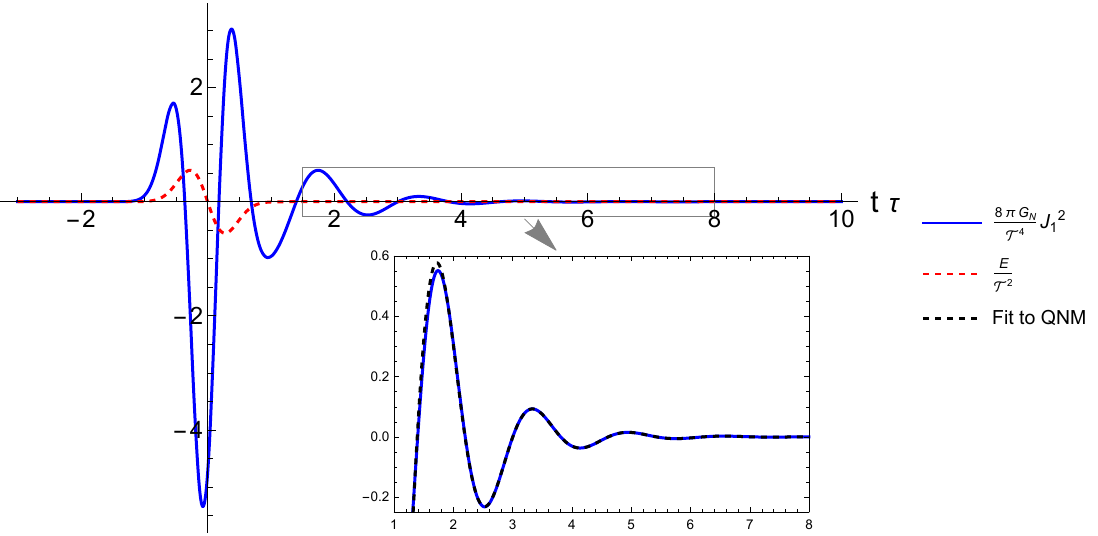}}
{\caption{ \label{F:currenteg} The expectation value of the current as a function of time (blue) in the presence of an external electric field (dashed red) for $\lambda=1/2$, $B=2\mathcal{T}^2$, and an electric field described by \eqref{E:dtGaussian} with $E_0 = 1/2 \mathcal{T}^2$ and $\tau \mathcal{T}=11/40$. The Fourier transform of the external electric field is supported at $1/\tau \sim \hbox{Re}(\omega_1)$ with $\omega_1$ the quasi normal mode (QNM) associated with the gauge field. The excitation of the quasi normal mode manifests in the boundary theory as a decaying oscillatory current.}}
\end{center}
\end{figure}

\begin{figure}[h]
\begin{center}
{\includegraphics[scale=0.8]{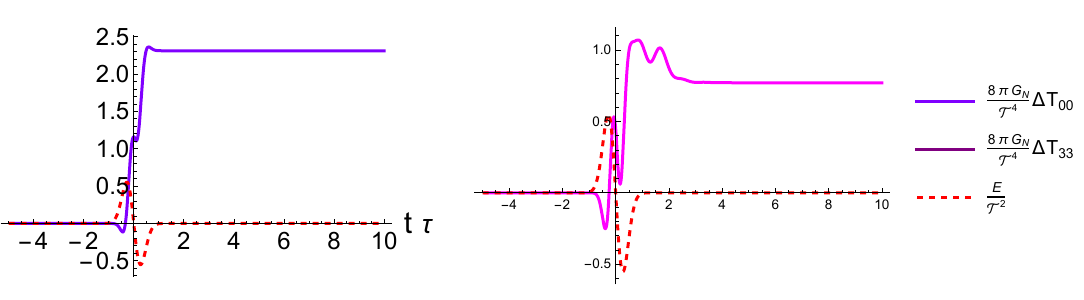}}
{\caption{ \label{F:metriceg} We plot components of the stress tensor  (purple and pink) relative to the stress tensor for an unpertrubed magnetic black brane at $\lambda=1/2$, $B=2\mathcal{T}^2$, and an electric field described by \eqref{E:dtGaussian} with $E_0 = 1/2 \mathcal{T}^2$ and $\tau \mathcal{T}=11/40$ (dashed red). The stress tensor component $T_{33}$ exhibits non trivial dynamics at intermediate times, after the electric field has been turned off indicating non linear gravitational effects.}}
\end{center}
\end{figure}

If the value of the Chern-Simons coupling $\lambda$ is too small the quasi normal modes will not drift towards the real axis and we don't expect to see long lived oscillations of the current and stress tensor at late times. Nevertheless, we do expect to see a decaying excitation whose lifetime is proportional to the imaginary component of the quasi normal frequency.
To this end, we consider the response of the current and stress tensor an electric field of strength $E_0 = 0.5 \mathcal{T}^2$ and width $\tau \mathcal{T} = 11/40$ (with $\mathcal{T}$ as before), a magnetic field of $B_0 = 2 \mathcal{T}^2$ and an anomaly with $\lambda = 1/2$. For these parameters we find that the final temperature of the system, long after the electric field has been turned on, asymptotes to $T_f \sim 1.65 T_0$. In this case the lowest quasi normal mode of the late time solution is given by $\omega_1 = 3.90-1.14 i$, so that $Re(\omega_1) \sim 1.1/ \tau$. 
The behavior of the current in this setup is depicted in figure \ref{F:currenteg} and that of the stress tensor in figure \ref{F:metriceg}.
Here, while the current decays quickly to zero, the spatial components of the stress tensor remains excited at intermediate times $1 \lesssim t \lesssim 3$ after the electric field has been turned off. As before, this transient behavior involves strong gravitational effects where non linear gravity is important in capturing the correct dynamics.

\section{Discussion}
\label{S:discussion}
In this work we've studied quenches in magnetically charged anomalous thermal states. The momentum independent quasi normal modes associated with the dual black hole description of these thermal states exhibit long lived excitations which may be triggered by an appropriately tuned external electric field. Long lived excitations of this type have been previously observed in \cite{Ammon:2016fru,Ammon:2017ded,Haack:2018ztx} and it seems likely that they are generic, at least for theories with a holographic dual. In this context, there are several directions one may pursue. 

If the anomalous resonance effect is robust, it stands to reason that such an effect will manifest itself in other, non holographic, systems where the anomaly is sufficiently strong. In this case, the effect has the potential of being observed experimentally in systems whose effective field theory description involves chiral fermions, e.g., Weyl semi-metals.

On a somewhat different note, one may also attempt to observe the anomalous resonance effect in a fully controllable holographic dual pair (where the field theory description and the gravitational dual are well understood). As discussed earlier,  the action \ref{E:ansatz} is related to the STU ansatz for the $D=5$, $N=2$ gauged supergravity action \cite{Behrndt:1998jd}. One may check that the solution \eqref{E:chargedsol} with $q^i=0$ is a solution to the gauged supergravity action of \cite{Behrndt:1998jd} (first discussed in \cite{Donos:2011qt}) but quasi normal modes of these magnetically charged black brane solutions will neccessarily involve fluctuations of the scalar fields of the STU ansatz. 

Our expectation is that the quasi normal modes associated with the anomalous resonance effect computed here will be unmodified when including the aforementioned additional scalar fields. The reason being that the quasi normal modes of interest are vector modes which should decouple from the scalar quasi normal modes.

In parallel to the anomalous resonance effect the authors of \cite{Donos:2011qt} discussed spatially modulated instabilities of the magnetically charged black brane solution (see also, e.g., \cite{Nakamura:2009tf,Ooguri:2010kt} for a general discussion). In \cite{Donos:2011qt} it was argued that the magnetically charged black brane solution is unstable to a spatially modulated phase with the critical temperature, $T_c$, for this instability being rather low. In the presence of such a phase transition one might worry that the anomalous resonance effect is unobservable since the spatially modulated phase dominates the dynamics at low temperatures where the imaginary component of the quasi normal modes becomes sufficiently small. Of course, this is a matter of scales: whether the imaginary part of the quasi normal mode at $T>T_c$ is negligible. A full numerical analysis of the critical temperature and the associated quasi normal modes of the homogenous phase should be able to resolve this issue. 

Apart from the anomalous resonance effect, an additional effect associated with the anomaly, referred to as an anomalous trailing effect was discussed in \cite{Haack:2018ztx} and previously in \cite{Bu:2015ika,Bu:2015ame,Bu:2016vum,Bu:2018drd}. As it turns out, if the late time value of the (anomalous) background gauge field differs from its value at early times then the expectation value of the late time current will be sensitive to this (possibly gauge dependent) shift. The analysis of the anomalous trailing effect associated with the action \eqref{E:ansatz} is almost identical to the one provided in \cite{Haack:2018ztx} and will not be repeated here. We have checked, in several examples, that a sufficiently small Gaussian electric field perurbation indeed generates an anomalous trailing effect. A full study of anomalous trailing for strong background electric fields is left for future work.

\section*{Acknowledgements}
We would like to thank M. Ammon, O. Bergman, A. Buchel, S. Cremonini, J. Gauntlett, M. Haack, G. Lifschytz and C. Rosen for useful comments and feedback. We would also like to thank the organizers and participants of the KITP program "The many faces of relativistic fluid dynamics" and of the NORDITA program "Hydrodynamics at all scales" where this work was finalized. The authors are suppored in part by a binational science foundation grant and  by the National Science Foundation under Grants No. NSF PHY-1748958 and PHY-2309135.

\begin{appendix}
\section{Holographic prescription}
\label{A:prescription}

A holographic renormalization program for Einstein-Chern-Simons-Maxwell theory with a single $U(1)$ gauge field,
\begin{equation}
	S = \frac{1}{16\pi G_N} \int \sqrt{-g}\left(R + 12 - \frac{1}{4}F_{\mu\nu}F^{\mu\nu} - \frac{\kappa}{4} \epsilon^{\mu\nu\rho\sigma\tau}A_{\mu}F_{\nu\rho}F_{\sigma\tau} \right)d^5x
\end{equation}
was carried out in \cite{Sahoo:2010sp}. Writing the metric and gauge field in a Fefferman-Graham coordinate system
\begin{align}
\begin{split}
\label{E:FGcoordinates}
	g_{\mu\nu}dx^{\mu}dx^{\nu} &= \frac{d\chi^2}{4\chi^2} + \frac{1}{\chi} \left(g_{mn}^{(0)} + g_{mn}^{(2)}\chi + g_{mn}^{(4)}\chi^2 + h_{mn}^{(4)} \chi^2 \ln\chi + \ldots \right) dx^m dx^n\\
	A_{m} &= A_{m}^{(0)} + A_{m}^{(2)} \chi + B_{m}^{(2)} \chi\ln\chi + \ldots \\
	A_{\chi} & = 0\,,
\end{split}
\end{align}
the resulting one point function for the stress tensor and current are
\begin{align}
\begin{split}
\label{E:SYresult}
	8\pi G_N \langle T_{mn} \rangle &= 2 t_{mn} + (3+2\alpha_1) h_{mn}^{(4)} - (1-8\alpha_2) \frac{1}{4} \left(\frac{1}{4} {F^{(0)}}^2 g^{(0)}_{mn} + F_{ma}^{(0)} g^{(0)aa'}F^{(0)}{}_{a'n}\right) \\
	8 \pi G_N \langle J^{m} \rangle & = g^{(0)mn} \left(A_{n}^{(2)} + B_{n}^{(2)} (1+\alpha_1+8\alpha_2)\right) - \frac{\kappa}{2} \epsilon^{mnsd}A_{n}^{(0)}F_{sd}^{(0)} \,.
\end{split}
\end{align}
Here
\begin{multline}
	t^{mn} = g_{mn}^{(4)} 
	- \frac{1}{2} g^{(2)}_{ma}g^{(0)ab}g^{(2)}_{bn} 
	+ \frac{1}{4} g^{(0)ab}g_{(2)ba}g^{(2)}_{mn} 
	\\
	- \frac{1}{8}\left(\left(g^{(0)ab}g_{(2)ba}\right)^2 
	+ g^{(0)ab}g^{(2)}_{bc}g^{(0)cd}g^{(2)}_{d a}\right)g^{(0)}_{mn} 
	- \frac{1}{48}{F^{(0)}}^2 g_{mn}^{(0)} \,,
\end{multline}
$h^{(4)}_{mn}$, $B^{(2)}_m$ and $g^{(2)}_{mn}$ are determined by the equations of motion
and we have used $F^{(0)} = dA^{(0)}$, and ${F^{(0)}}^2 = F_{mn}^{(0)}g^{(0)mm'}g^{(0)nn'}F^{(0)m'n'}$.  In what follows we will raise and lower indices of the field strength $F^{(0)}_{mn}$ using the boundary metric, $g^{(0)mn}$, e.g., $F^{(0)m}{}_{n} = g^{(0)m'm} F^{(0)}_{m'n}$. The coefficients $\alpha_i$  are scheme dependent undetermined coefficients associated with finite counterterms which can be added in the  holographic renormalization program associated with the trace anomaly and the field strength squared \cite{Sahoo:2010sp}. There are other counterterms which can be added which will modify the expectation value of the stress tensor in a non flat background metric $g^{(0)}_{mn}$ \cite{Buchel:2012gw}. We have refrained from writing these since we will be setting the boundary metric to the Minkowski metric  so they won't be relevant to our discussion.

We point out that when the background metric is flat the aforementioned counterterms are proportional to each other and vanish whenever $\alpha_1+8\alpha_2=0$. To see this we note that choosing $g^{(0)}_{mn} = \eta_{mn}$ considerably simplifies the expressions for $h^{(4)}_{mn}$, $B^{(2)}_m$ and $g^{(2)}_{mn}$ which now take the form
\begin{align}
\begin{split}
	h^{(4)}_{mn} & = \frac{1}{8} \left(\frac{1}{4} {F^{(0)}}^2 g^{(0)}_{mn} + F_{ma}^{(0)}  F^{(0)a}{}_{n}\right)  \\
	B^{(2)}_{m} & = -\frac{1}{4} \nabla_n F^{(0)n}{}_{m}\\
	g^{(2)}_{mn} & = 0\,,
\end{split}
\end{align}
where $\nabla$ is a covariant derivative associated with the boundary metric $g^{(0)}_{mn}$.
Inserting these expressions into \eqref{E:SYresult} brings the stress tensor and current into the form
\begin{align}
\begin{split}
\label{E:SYresultsimple}
	8\pi G_N \langle T_{mn} \rangle &= 2 g^{(4)}_{mn} - \frac{2}{48}{F^{(0)}}^2 g_{mn}^{(0)}  +  \frac{1}{4} \left(\frac{1}{2} + \alpha_1 + 8\alpha_2 \right) \left(\frac{1}{4} {F^{(0)}}^2 g^{(0)}_{mn} + F_{ma}^{(0)} F^{(0)a}{}_{n}\right) \\
	8 \pi G_N \langle J^{m} \rangle & = \left(g^{(0)mn}  A_{n}^{(2)} -\frac{1}{4}\nabla_n  F^{(0)nm}(1+\alpha_1+8\alpha_2)\right) - \frac{\kappa}{2} \epsilon^{mnsd}A_{n}^{(0)}F_{sd}^{(0)} \,.
\end{split}
\end{align}


Energy momentum and charge conservation of the stress tensor and current in \eqref{E:SYresult} read
\begin{align}
\begin{split}
\label{E:holographicC}
	8 \pi G_N \nabla_{m} \langle T^{mn} \rangle &= F^{(0)nm} \left(8 \pi G_5 \langle J_{m} \rangle + \left(\frac{1}{2} + 2\alpha_1+16  \alpha_2\right) \frac{1}{4}\nabla_n F^{(0)nm} + \frac{1}{2} \kappa \epsilon_{m}{}^{rst}A^{(0)}_{r}F^{(0)}_{st}  \right) \\
	8 \pi G_N \nabla_{m} \langle J^{m} \rangle & =  \frac{\kappa}{8} \epsilon^{mnrs}F^{(0)}_{mn}F^{(0)}_{rs} \,,
\end{split}
\end{align}
where indices have been raised with the (inverse) boundary metric $g^{(0)mn}$.
The first term on the right hand side of \eqref{E:holographicC} is the standard Joule heating term associated with an external electromagnetic field $F^{(0)\mu\nu}$ and a current $J_m$. The second term on the right of \eqref{E:holographicC} can be thought of as a Joule heating term associated with a conserved current proportional to $\nabla_m F^{(0)mn}$. The last term on the right of \eqref{E:holographicC} indicates that the consistent current $J^{\mu}$ is not gauge invariant \cite{BARDEEN1984421}.

If we want to obtain a canonical Joule heating term on the right hand side of the energy momentum conservation equation \eqref{E:holographicC} to be associated with a single covariant current, we can define
\begin{equation}
	8 \pi G_5 J_{cov}^{m} = 8 \pi G_5  J^{m} + \left(\frac{1}{2} + 2\alpha_1+16  \alpha_2\right) \frac{1}{4}\nabla_n F^{(0)nm}  + \frac{\kappa}{2} \epsilon^{mrst} A_{r}^{(0)} F^{(0)}_{st}
\end{equation}
such that the conservation equations take the form
\begin{align}
\begin{split}
\label{E:holographicCcov}
	\nabla_{\mu} \langle T^{mn} \rangle &= F^{(0)nm}  \langle J_{cov\,m} \rangle \\
	8 \pi G_N \nabla_{m} \langle J_{cov}^{m} \rangle & =  \frac{3\kappa}{8} \epsilon^{mnrs}F^{(0)}_{mn}F^{(0)}_{rs} \,.
\end{split}
\end{align}
We point out that while the unique covariant current $J_{cov}^m$ is gauge invariant and satisfies the canonical Ward identity associated with a Joule heating term, the current $J_m$ is what will be evaluated by, say, a diagrammatic evaluation of the $U(1)$ current associated with the symmetry of the problem.

In our setup there are three $U(1)$ fields and a mixed Chern-Simons term. The resulting one point function can be deduced from \eqref{E:SYresultsimple} via symmetry. We find
\begin{align}
\begin{split}
\label{E:holographicdictionary} 
	8 \pi G_N \langle T_{mn} \rangle &= 2 g_{mn}^{(4)} - \frac{1}{24}\sum_{i=1}^{3}{F_i^{(0)}}^2 g^{(0)}_{mn}  +\frac{1}{8}\left(1+2 \alpha_1+16 \alpha_2\right)\sum_i \left(\frac{1}{4} {F_i^{(0)}}^2 g^{(0)}_{mn} + F_{i\,ma}^{(0)} F_i^{(0)a}{}_{n} \right)\\
	8 \pi G_N \langle J_{i\,cov}^{m} \rangle & = g^{(0)mn}  A_{i\,n}^{(2)} - \frac{1}{8}\left(1-2 \alpha_1-16 \alpha_2\right) \nabla_{n}F^{(0)\,nm}_{i}  \,.
\end{split}
\end{align}

The metric and gauge field we are interested in are given by the Eddington-Finkelstein coordinate system \eqref{E:Sansatz} and not the Fefferman-Graham coordinate system of \eqref{E:holographicdictionary}. In order to relate the asymptotic coefficients, $A^{(0)}_{m}$, $A^{(2)}_{m}$, $B^{(2)}_{m}$ and $g^{(4)}_{mn}$ appearing in \eqref{E:holographicdictionary} to the expansion in \eqref{E:BV} we need to carry out a coordinate transformation from \eqref{E:Sansatz} to  \eqref{E:FGcoordinates}. 
Recall that \eqref{E:BV} reads
\begin{align}
\begin{split}
\label{E:BV2}
	r_*^{-1} a & = a^{(0)}(t_*) + \frac{\partial_{t_*} a^{(0)}(t_*)}{\rho_*} - \frac{\partial_{t_*}^2 a^{(0)}}{2 \rho_*^2}\ln {\rho_*} + \frac{a^{(2)}(t_*) }{\rho_*^2} + \mathcal{O}(\rho_*^{-3}) \\
	r_*^{-1} \tilde{a} & = \frac{c+4 \beta_* \lambda a^{(0)}(t_*)}{\rho_*^2} + \mathcal{O}(\rho_*^{-3})\\
	r_*^{-1} \Sigma &= \rho_* + \mathcal{O}(\rho_*^{-3}) \\
	g & = 1 - \frac{\partial_{t_*} a^{(0)}(t_*)}{12 \rho_*^4} \ln \rho_* + \frac{g^{(4)}}{\rho_*^4} + \mathcal{O}(\rho_*^{-5})\\
	r_*^{-2} A & = \frac{\rho_*^2}{2} - \frac{3 \beta_*^2 + 2 (\partial_{t_*} a^{(0)}(t_*))^2}{12 \rho_*^2} \ln \rho_* + \frac{A^{(2)}}{\rho_*^2} + \mathcal{O}(\rho_*^{-3}) \\
\end{split}
\end{align}
(with $t_* = t r_*$, $\beta_* = B r_*^{-2}$, $\rho_* = r r_*^{-1}$) once we set $r_* = r_h$. For pedagogical reasons we have used here \eqref{E:BV2} which is a slight generalization of \eqref{E:BV}.
While it is practically impossible to find a closed form expression for a coordinate transformation that will take is from \eqref{E:Sansatz}  to \eqref{E:holographicdictionary}, it is straightforward to compute its near boundary series expansion. 

The coordinate transformation
\begin{align}
\begin{split}
\label{E:CT}
	r &= \frac{1}{\sqrt{\chi}} + \mathcal{O}(\chi^{\frac{3}{2}})\,, \qquad
	t = x^0 -  \sqrt{\chi} + \mathcal{O}(\chi^{\frac{5}{2}})\,, \\
	x & = x^1\,, \qquad
	y = x^2\,, \qquad
	z  = x^3\,, 
\end{split}
\end{align}
takes us from the coordinate system \eqref{E:Sansatz} to \eqref{E:FGcoordinates}. 
The resulting boundary metric is given by
\begin{equation}
	g^{(0)}_{mn}  = \eta_{mn}\,.
\end{equation}
Similarly, using \eqref{E:CT} to transform the expansion \eqref{E:BV} into the variables in \eqref{E:FGcoordinates} we find
\begin{subequations}
\label{E:EFToFGconversion}
\begin{align}
\begin{split}
	r_*^{-1} A_1^{(0)} &=  a^{(0)}(x_*^0) dx^2 + \beta_* r_* x^2 dx^3 \\
	r_*^{-1} A_2^{(0)} &=  a^{(0)}(x_*^0) dx + \beta_* r_*  x^3 dx^1 \\
	r_*^{-1} A_3^{(0)} &= \beta_* r_*  x^1 dx^2\,
\end{split}
\end{align}
with $x_*^0 = x_0 r_*$
for the boundary values of the gauge field.
The near boundary values of the gauge field are given by
\begin{align}
\begin{split}
	A_1^{(2)} & = r_*^3 \left(a^{(2)}(x^0_*) - \frac{1}{2} \left(1-\ln r_*\right) \partial_{x_*^0} a^{(0)}(x_*^0)  \right) dx^2\,, \\
	A_2^{(2)} & = r_*^3 \left(a^{(2)}(x^0_*) - \frac{1}{2}  \left(1-\ln r_*\right)  \partial_{x_*^0} a^{(0)}(x_*^0)  \right)  dx^1\,, \\
	A_3^{(3)} & = r_*^3\left(c + 4  \beta_* \lambda a^{(0)}(x_*^0) \right)dx^0\,,
\end{split}
\end{align}
The near boundary value of the metric reads
\begin{align}
\begin{split}
	g^{(4)}_{00} &= -\frac{r_*^4}{96}\left(144 A^{(2)}(x_*^0) +\left(1+12 \ln r_*\right) \left(3 \beta_*^2 + 2 \left(\partial_{x^0_*} a^{(0)}(x_*^0)\right)^2 \right) \right) \\
	g^{(4)}_{11} & = g^{(4)}_{22} =  \frac{r_*^4}{288}\left(- 144 A^{(2)}(x_*^0) +288 g^{(4)}(x_*^0) + 9 \left(1-4 \ln r_*\right) \beta_*^2 - 10 \left(\partial_{x^0_*} a^{(0)}(x^0_*)\right)^2  \right)  \\
	g^{(4)}_{33} &= \frac{r_*^4}{288}\left(- 144 A^{(2)}(x_*^0) -576  g^{(4)}(x_*^0) + 9  \left(1-4 \ln r_*\right) \beta_*^2  - 2(5+36 \ln r_*) \left(\partial_{x^0_*} a^{(0)}(x_*^0)\right)^2 \right)  \,,
\end{split}
\end{align}
\end{subequations}
(and the other components of $g^{(4)}_{mn}$ vanish).

Inserting \eqref{E:EFToFGconversion} into \eqref{E:holographicdictionary} 
and writing
\begin{equation}
	2 \ln r_c = \alpha_1+8 \alpha_2\,.
\end{equation}
we find,
\begin{subequations}
\label{E:duality}
\begin{align}
\begin{split}
	8 \pi G_5 \langle J_{1\,cov}^m \rangle &= \delta^m_2 r_*^3 \left( a^{(2)} (x_*^0) + \frac{1}{8}  \left(-3 +4 \ln \frac{r_*}{r_c} \right)  \partial_{x_*^0}^2 a^{(0)}(x_*^0) \right)\\
	8 \pi G_5 \langle J_{2\,cov}^m \rangle &= \delta^m_1 r_*^3 \left(  a^{(2)} ( x^0) + \frac{1}{8}   \left(-3 +4  \ln \frac{r_*}{r_c} \right)  \partial_{x_*^0}^2 a^{(0)}( x_*^0) \right) \\
	8 \pi G_5 \langle J_{2\,cov}^m \rangle & = - \delta^m_0 r_*^3 \left(c+ 4 \beta_* \lambda a^{(0)}( x_*^0)\right)
\end{split}
\end{align}
for the expectation value of the current and
\begin{align}
\begin{split}
	16 \pi G_5 \langle T_{00} \rangle &= r_*^4\left( -6 A^{(2)}(x_*^0) 
		-\frac{1}{2} \left(\frac{1}{3} +2 \ln \frac{r_*}{r_c} \right) \left( \partial_{x_*^0} a^{(0)}(x_*^0) \right)^2 
		+ \frac{3}{4} \left(1 - 2 \ln \frac{r_*}{r_c} \right) \beta_*^2  \right)\\
	16 \pi G_5 \langle T_{11} \rangle &  =r_*^4 \left( -2 A^{(2)}(x_*^0) 
		+ 4 g^{(4)}(x_*^0) 
		+ \frac{7}{36}\left( \partial_{x_*^0} a^{(0)}(x_*^0) \right)^2  
		-\frac{1}{4} \left(1+2 \ln \frac{r_*}{r_c} \right)\beta_*^2 \right) \\
	16 \pi G_5 \langle T_{33} \rangle &= r_*^4\left( -2 A^{(2)}(x_*^0) - 8 g^{(4)} (x_*^0)
		+ \frac{1}{2}\left(\frac{8}{9} -2  \ln \frac{r_*}{r_c} \right) \left(  \partial_{x_*^0} a^{(0)}(x_*^0) \right)^2  
		-\frac{1}{4} \left(1+2 \ln \frac{r_*}{r_c} \right)\beta_*^2 \right)  \\
	\end{split}
\end{align}
\end{subequations}
and $\langle T_{22} \rangle= \langle T_{11} \rangle$
for the stress tensor.

As an example, consider the uncharged magnetic black brane solution \eqref{E:chargedsol}. Expanding \eqref{E:chargedsol} as in \eqref{E:BV}  and inserting this into \eqref{E:duality}, we find
\begin{align}
\begin{split}
\label{E:TEV}
	16 \pi G_5 \langle T^{00} \rangle &=  3 \left( r_{h}^4 +\frac{B^2}{4} \left(1 - 2 \ln \left(\frac{r_h}{r_c}\right)\right) \right)  \\
	16 \pi G_5 \langle T^{11} \rangle & = \left( r_h^4 +\frac{B^2}{4} \left(-1 - 2 \ln \left(\frac{r_h}{r_c}\right)\right)\right) \,.
\end{split}
\end{align}
The scale $r_*$ which we have used in intermediate computations has dropped out as expected. The logarithm appearing in the expectation value of the stress tensor in \eqref{E:TEV} is scheme dependent. 

When constructing numerical solutions to the equations of motion we have used $r_*=r_h$ the initial temperature of the magnetic black brane. We have also chosen $r_c=r_h$ to simplify the resulting stress tensor and current, viz., \eqref{E:EVcurrents} and \eqref{E:EVstress}.

\end{appendix}

\bibliographystyle{JHEP}
\bibliography{magneticBH}

\end{document}